\documentclass[aps,twocolumn,amsmath,amssymb]{revtex4-1}
\usepackage{graphicx}% Include figure files
\usepackage{dcolumn}% Align table columns on decimal point
\usepackage{bm}% bold math

\usepackage{epstopdf} 
\begin{document}

\title{Fabrication of Electronic Fabry-Perot Interferometer in the Quantum Hall Regime}

\author{S. An$^{1}$, S. Glinskis$^{1}$, W. Kang$^{1}$, L. E. Ocola$^{2}$, L. N. Pfeiffer$^{3}$, K. W. West$^{3}$, K. B. Baldwin$^{3}$}
\affiliation{$^1$James Franck Institute and Department of Physics, University of Chicago, Chicago, IL 60637\\
$^2$Center for Nanoscale Materials, Argonne National Laboratory, Argonne, IL 60439\\
$^3$Department of Electrical Engineering, Princeton University, Princeton NJ 08544}

\begin{abstract}
A fabrication method for electronic quantum Hall Fabry-Perot interferometers (FPI) is presented. Our method uses a combination of e-beam lithography and low-damage dry-etching to produce the FPIs and minimize the excitation of charged traps or deposition of impurities near the device. Optimization of the quantum point contacts (QPC) is achieved via systematically varying the etch depth and monitoring the device resistance between segmented etching sessions. The etching is stopped when a desired value of resistance is obtained. Finer control of interference trajectories is obtained by the gate metallized inside the etched area by e-beam evaporation. Our approach allows for a control of the delicate potential bending near the quantum well by tuning the confining potential of the quantum point contacts. 
\end{abstract}

\maketitle

Topological qubits have been proposed to realize a fault-tolerant quantum computer which is protected and immune from the effects of decoherence \cite{DasSarma05,Nayak08}. 
Such an approach to quantum computation based on topological qubits is predicated on the existence of non-Abelian anyons, which are elementary excitations postulated to exist in the $\nu = 5/2$ fractional quantum Hall (FQH) state\cite{Moore91}. While the non-Abelian anyons are highly anticipated, reports of experimental detection have been controversial \cite{Willett09,Willett10,An11}.

Interference experiments involving electronic Fabry-Perot interferometer (FPI) in quantum Hall system has been proposed to  to detect non-Abelian anyons \cite{Stern06,Bonderson06}. In experiments carried out so far, an electronic FPI is realized by applying bias voltages to gate electrodes patterned on the surface of high mobility GaAs/AlGaAs heterostructures\cite{Willett09,Willett10,Zhang09,Ofek10}. The depletion of electrons in the underlying quantum well by the negative bias voltages from the surface in principle provides a simple way to define the required interferometer geometry. However, application of bias voltages can also excite spurious charge traps and induce scattering sources, degrading mobility and coherent length of the current-carrying excitations in delicate quantum Hall states.

In this paper we present an alternative fabrication process for FPI devices made using dry-etching to define the potential profile of the interferometer. We have also developed a method to optimize the size of the interference signal through controlling the amplitude of backscattering in the FPIs. The advantage of such approach is elimination of the use of bias voltages to minimize the excitation of spurious charges and reduction of the parameter space for detection of interference signals. 

\begin{figure}
\includegraphics[width=3.25in]{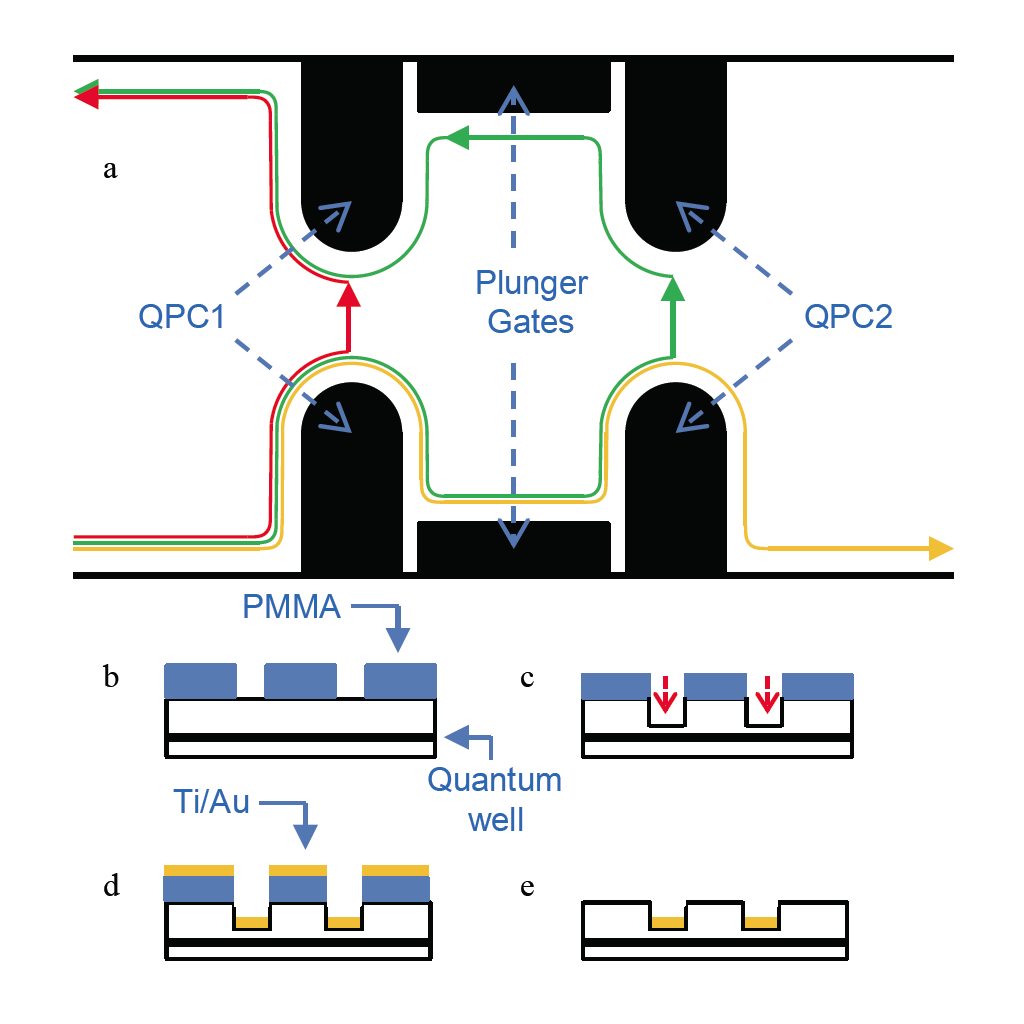}
\caption{\label{fig:fig1} Fig. 1. Schematic showing the conceptual overview of FPI device (a) and processing steps for creating FPI device on GaAs/AlGaAs quantum well structure (b-e): (b) creating device pattern in PMMA by e-beam lithography; (c) ICP dry etching; (d) evaporating Ti/Au onto the etched device pattern; (e) self-aligned gate metals inside the etched trenches after lift-off. } 
\end{figure}

Our design of the FPI devices consists of two quantum point contacts (QPCs) and and two plunger gates in a Hall-bar system to produce interfering particle streams optimized for analysis of quantum statistics of edge states. The two QPCs, which individually create a narrow constriction where current-carrying particles can backscatter, behave as a beam-splitter that plays in the quantum interference. Figure 1 (a) illustrates conceptual view of such a FPI device with two QPCs and two plunger gates. In quantum Hall system, the currents flow chirally along the edges of two-dimensional electrons. Near a narrow constriction formed by a QPC, some current-carrying particles are able to tunnel across to the opposite edge, resulting in part of the incoming current being backscattered. The process is similar to an optical beam splitter partially transmitting and reflecting light. The presence of two such constrictions in series generates two streams of backscattered particles that interfere with each other, making this device an analogue of optical FPI with two beam splitters.

The path difference of two interfering particles ends up encircling (braiding) the inner area 
and the localized particles in the central part of the device. Such particle braiding is topologically equivalent to exchanging the particles twice. The quantum interference produced by electronic FPI not only reflects the Aharanov-Bohm phase acquired by the braiding particle in the magnetic field, but also the nature of quantum exchange (braiding) statistics of the interfering particles.
We chose circular device and the symmetrically designed QPCs with sharp tips. We didnÕt consider wet etching processes in defining the device geometries because of difficulty in dimensional controls from lateral etching and uneven floor near the corners and edges of sub-micron features as well as possible deposition of charged ions on etched surfaces.

\begin{figure}
\includegraphics[width=3.25in]{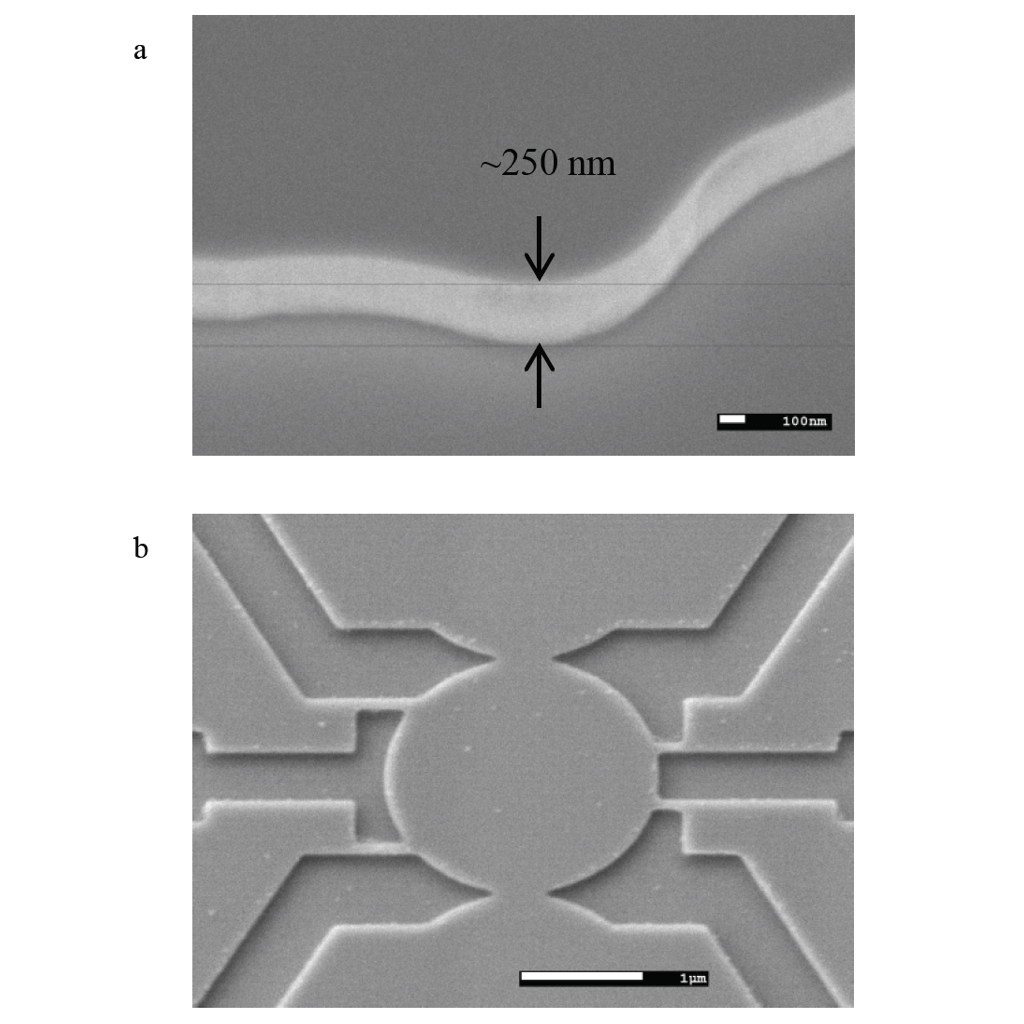}
\caption{\label{fig:fig2} Fig. 2. SEM images of dry-etched patterns with no metals deposited: (a) test pattern with etched depth of  250 nm; (b) FPI device with etched depth of  70 nm. } 
\end{figure}

The fabrication method presented here uses e-beam lithography and dry-etching to produce the device geometry and minimize the excitation of charged traps or deposition of impurities near the device. Further control of QPC properties or modulation of quasiparticle paths are provided by the gate metallization deposited inside the etched area by e-beam evaporation. The key issue is minimize any degradation of electron mobility near the device area to keep the coherence length of the encircling edge currents as long as possible. Fig. 1 (b)-(e) show the overview of fabrication procedures.
The FPI devices were fabricated on a symmetrically doped GaAs/AlGaAs quantum well with a ultrahigh mobility of $2.7 \times $10$^{7}$ cm$^2$/Vs and electron density of 3.5$\times$10$^{11}$ cm$^{-2}$ (both at cryogenic temperatures) grown by molecular beam epitaxy. The 30 nm wide quantum well is located about 100 nm below the sample surface, with symmetrical Si doping layers below and above the quantum well. 
For the patterning of FPI device, PMMA was spin-coated and exposed using a Jeol 9300FS e-beam lithography system. 
Proximity effect corrections were made on the device pattern. 

\begin{figure}
\includegraphics[width=3.25in]{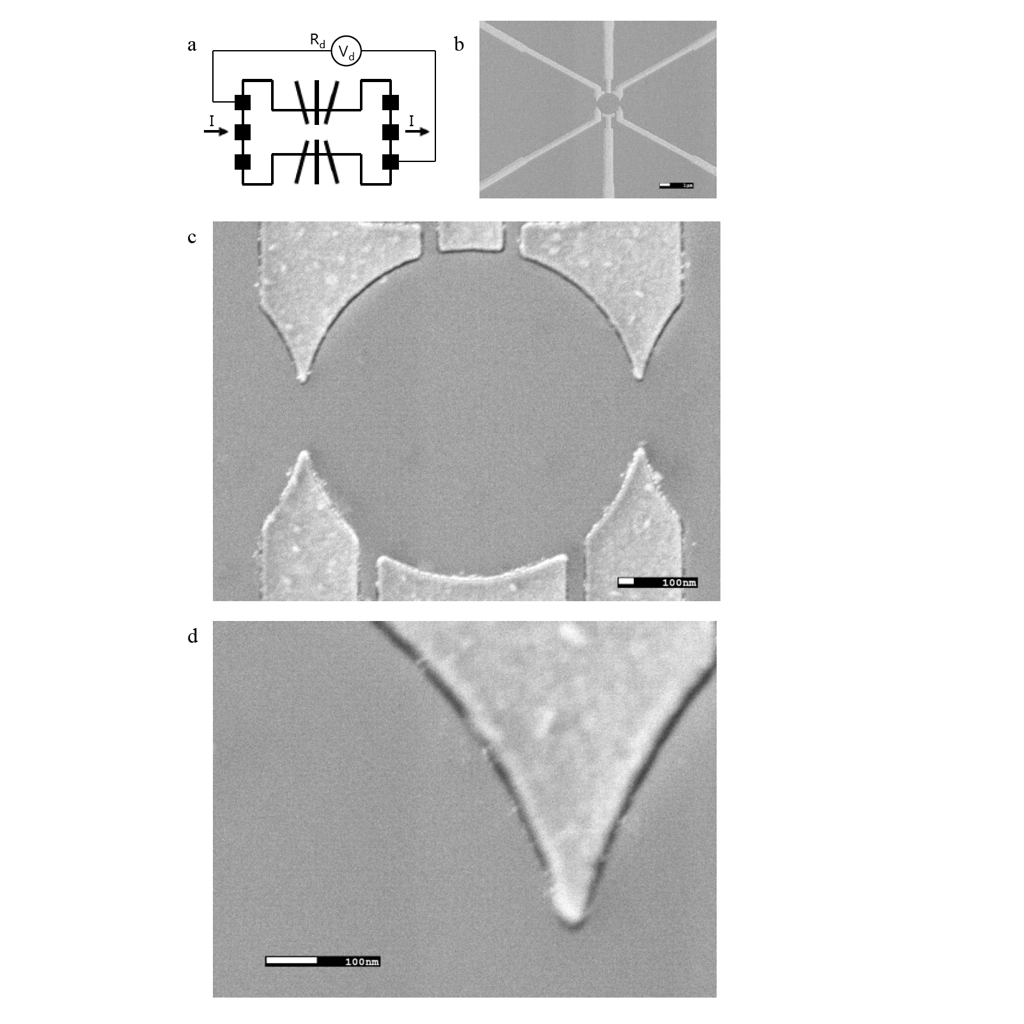}
\caption{\label{fig:fig3} Fig. 3. Fabricated FPI device: (a) diagram of wet etched Hall bar, ohmic contacts, and fabricated FPI; (b-d) SEM images of fabricated FPI with subsequent zoom-ins showing device geometry and one of the QPC tips.}\end{figure}

The pattern was dry etched using an inductively coupled plasma (ICP) etcher (RIE Oxford PlasmaLab 100 system) in a Cl$_2$/BCl$_2$/Ar gas mixture. This recipe produced vertical, clean trenches in GaAs with relatively slow etch rate of 70 - 100 nm/min for ease of depth control and with very little lateral etch. Fig. 2 shows tilted view SEM images of etched GaAs/AlGaAs structure with this recipe. Etching was performed in a series of several etching sessions with duration of 10 - 30 sec each, with the sample taken out in between the sessions and the device resistance measured using a probe station at room temperature, to ensure the QPCs are neither insulating (etched too deep) nor too conductive (etched too shallow). Target values for device resistance at the end of the etching procedure was determined by its correlations with device resistance at cryogenic temperatures observed in the previous FPI devices fabricated from the same wafer.

The FPI resistance at room temperature was monitored as a function of accumulated etched time, where a steep increase is observed as the etched surface approaches the quantum well layer. Failure of stopping the etch within the target range at 300 K usually results in FPIs that are often insulating at cryogenic temperatures. The resistance was measured using a probe station under microscope, with the PMMA layer still in place, by probes touching the ohmic contacts. 
The self-aligned Ti/Au gate was metallized via e-beam evaporation within the etched trenches.
Lift-off was enhanced by the gate metals sitting within the trenched area well below the PMMA layer.

Fig. 3 shows SEM micrographs of a FPI device along with a schematic of the measurement geometry. The device was made with two symmetrical QPCs with a width of 400 nm and two plunger gates of differing widths to provide more variety of area modulation. 
The diameter of the active region is roughly 2.2 $\mu$m, and the gap between the QPC and plunger gates is $\sim$125 nm. Self-aligned gate metals inside etched trenches were positioned very close to the pattern edges, showing nearly no sign of lateral etch.

We have reported the fabrication process of electronic Fabry-Perot interferometer in high mobility GaAs/AlGaAs quantum well structure using electron beam lithography and ICP dry etching. The steps outlined here allow for some control of the confining potential of the QPCs to detect quantum interference at low temperatures. Optimization of the FPI properties can be achieved by the feedback based on the measured device resistance as a function of the etched depth. Various processing features can be tweaked to achieve a desired value of FPI resistance. 

S.A., S.G., and W.K. acknowledge support by Microsoft Project Q, John Templeton Foundation, and NSF MRSEC Program through the University of Chicago Materials Center (DMR-0820054).  L.N.P, K.W.W., and K.W.B acknowledge partial support from the Gordon and Betty Moore Foundation and the NSF MRSEC Program through the Princeton Center for Complex Materials (DMR-0819860).\\

\end{document}